\documentclass[12pt,bezier]{article}
\usepackage{amsmath,amssymb,amsfonts,euscript}
\usepackage[usenames]{color}
\usepackage{graphicx}
\newcommand{\be}{\begin{equation}}
\newcommand{\ee}{\end{equation}}
\newtheorem{theorem}{Theorem}[section]
\newtheorem{lemma}[theorem]{Lemma}

\newtheorem{example}[theorem]{Example}
\newtheorem{definition}{Definition}[section]
\newtheorem{remark}[theorem]{Remark}
\newtheorem{proposition}[theorem]{Proposition}

\title{\bf\Large Modeling credit default swap premiums with stochastic recovery rate}
\vspace{-1cm}
\author
{ Z.Sokoot\thanks{Allameh Tabataba'i University, Tehran, Iran, E-mail: sokoot921@atu.ac.ir}\,\, ,\, N.Modarresi\thanks{Faculty of
Mathematics and Computer Science, Allameh Tabatabae'i University, Tehran, Iran. E-mail: n.modarresi@atu.ac.ir.} ,\,\, F.Niknejad \thanks{ Allameh
Tabataba'i University, Tehran, Iran, E-mail: f.niknejad@atu.ac.ir.} }
\date{}

\numberwithin{equation}{section}
\begin{document}
\maketitle
\begin{abstract}
\noindent There are many studies on development of models for analyzing some derivatives such as credit default swaps (CDS). A continuous-time autoregressive
moving average (CARMA) driven by L\'evy process is applied for modeling the CDS premia. It is based on the stochastic recovery rate which is
time-varying during the maturity stage and makes the model suitable for evaluating the premium leg. We show that this model is a class of affine term
structure model.
By simulating the CARMA(2,1) process the effectiveness of this model in determining the most appropriate parameters are illustrated. Also, a real data set of daily CDS premia of some companies is used and a comparison of the Bayesian information criterion between the models is given.\\ \\
{\it Keywords\,\,\,\, CDS spread. Stochastic recovery rate. CARMA model}\\ \\
{\it JEL Classification\,\,\,\, C15. C32. G13}
\end{abstract}

\section{Introduction}
In financial markets, derivatives enable parties to trade specific financial risks and help to improve market efficiencies.
Derivatives are financial contracts which derive their value of a spot price time-series such as forwards, futures, options and swaps. One of the most
important and applicable derivatives is a credit default swap (CDS).
There are many studies about the valuation, hedging and modeling of credit risk \cite{bielechi}. As with any swap, valuing CDS involves calculating the
present value of two legs of the transaction, premium leg and default leg. Wemmenhove described a model for CDS spread form and presented a forecasting
model for it \cite{Wemmenhove}. Also, the determinants at CDS spread of European credit derivatives are analyzed \cite{jakovlev}.
A Markov model with the bankruptcy process following a discrete state space Markov chain in credit ratings are provided and the parameters of this
process are estimated using the observable data \cite{jarrow}.
The recovery rate and default probability have an important role in bond prices and there are many studies that attempted to model recovery and
comprehend their impact on depth values \cite{bakshi}. Also, modeling and empirically investigating the relationship between the components of
default risk as the probability of default and recovery rate are presented \cite{Altman}.\\
Jaskowaki et al. proposed a parsimonious reduced form continuous-time model that estimates expected recovery rates and the parameters were estimated by
using a Bayesian MCMC algorithm \cite{Jaskowaki}.
A class of models which has proven to be remarkable flexible structure for examining the dynamics of default-risk free bonds is an affine term
structure (ATS) model. This model suggests that interest rates at any point in time are a time-invariant linear function of a small set of common
factors. ATS models start from the assumption that there are no arbitrage opportunities in financial markets and implies the existence of a positive
stochastic process that prices the assets \cite{Duffie}. These models have been shown to work well in approximating true yield dynamics. \\
The statistical models for continuous-time dynamics based on CDS premiums of different reference entities are presented  \cite{Eifert}. The sampled
reversion of L\'evy-driven continuous-time ARMA (CARMA) process is used as an appropriate model for modeling the CDS premium leg. Brockwell et al.
introduced the CARMA processes and estimated the parameters of the model based on discrete form specially in equally spaced samples
\cite{ferrazzano}, \cite{Yang}, \cite{davis}.\\
A new representation of the calculated quasi-likelihood to compute maximum Gaussian likelihood for estimating the
parameters of time series with irregularly spaced data are presented \cite{Pham}.
For L\'evy-driven CARMA processes, estimation procedures which take into account the generally non-Gaussian nature of the measure are less well
developed.\\
Recovery rates play a critical role in the estimation and pricing of credit risk derivatives and it is often assumed to be constant and independent of
default but it is not realistic and fair. It is shown that the recovery rates can be volatile and moreover dependent to default intensity.
In order to different risks in a company and determining the default leg, we consider the stochastic recovery rate model and compare the results. We
provide the new model under term structure time series model and the fair default payment which is applied in a risky companies contract are
considered. We study on a certain class of background L\'evy CARMA process for modeling the CDS premium with stochastic recovery rate (CDSP-SRR). This
process is a stationary solution of a stochastic differential equation, so for the first step in real data analysis we use a test for stationarity and
then we apply a method for finding the best model for CDSP-SRR. Selecting appropriate order for CARMA(p,q) process in CDSP-SRR model is the same as
modeling the CDS premium with constant recovery rate (CDSP-CRR) \cite{Eifert}.
Also, the results of data analysis for some companies show that the distributions of jumps can be compound Poisson, Normal Inverse Gaussian (NIG) and
other L\'evy processes.\\
The rest of the paper is organized as follows. Section 2 is devoted to the preliminaries and CARMA models driven by L\'evy processes, ATS models and
basic definitions of CDS contracts . We show that the L\'evy-driven CARMA model is ATS, in section 3. In section 4, the dynamic of CDSP-SRR is
presented. Also, in section 5, we simulate the CDSP-SRR model and the certain CARMA process. The developed model is illustrated and compared with
CDSP-CRR. For comparing the estimated models we use the Bayesian information criteria (BIC) and chose the model giving the smallest BIC over the whole
set of candidates. We use a data set of daily CDS premia across 242 firms from the European and North American markets.

\section{preliminaries}
In this section we present the linear stationary CARMA model which is driven by L\'evy process. Then we provide descriptions of some concepts as ATS
model and CDS.

\subsection{L\'evy-driven CARMA model}\label{sub2.1}
We discuss on L\'evy-driven CARMA processes and review the definition and properties of them \cite{Yang}.
\begin{definition}
A second-order L\'evy-driven continuous-time ARMA (p,q) process is defined in terms of the following state-space representation of the formal equation.
For $t\geqslant0$
\vspace{-2mm}
\begin{align}
a(D)Y_{t}=b(D)DL_{t}
\end{align}
where D denotes differential with respect to t and $\lbrace L_{t}, t\geqslant 0\rbrace$ is a L\'evy process with finite second moments, and
\begin{align}
a(z)& := z^{p} + a_{1}z^{p-1} + ...+ a_{p},\label{2.2}\\
b(z)& := b_{0} + b_{1}z + ...+ b_{p-1}z^{p-1}\label{2.3}
\end{align}
where the coefficients  $a_{1},...,a_{p},b_{0},...,b_{p-1}$  are complex-valued coefficients such that \\$b_q=1$ and $ b_{j} = 0 $ for $q < j < p $.
\end{definition}
To avoid trivial and easily eliminated complications, we assume that a(z) and b(z) have no common factors. The state-space representation consists of
the observation and state equations,
\begin{align}
Y_{t}={\bf b}^{\prime}X_{t}\label{2.4}\\
dX_{t}-\boldsymbol AX_{t}dt=\boldsymbol edL_{t},\label{2.5}
\end{align}
where $\{X_t, t\geqslant0\}$ is a state process which is the solution of the equation and
\begin{align}
\boldsymbol A:=
\begin{bmatrix}
0 &1 &0 &\dots &0\\
0 &0 &1&\dots &0\\
\vdots &\vdots &\vdots &\ddots & \vdots \\
0 &0 &0&\dots &1\\
-a_{p} &-a_{p-1} &-a_{p-2}&\dots &-a_{1}\\
\end{bmatrix}
,\, \, \, \, \, \
\boldsymbol e:=
\begin{bmatrix}
0 \\
0 \\
\vdots \\
0 \\
1\\
\end{bmatrix}
,\, \, \, \, \, \
\boldsymbol b:=
\begin{bmatrix}
b_{0} \\
b_{1} \\
\vdots \\
b_{p-2} \\
b_{p-1}\\
\end{bmatrix}
\end{align}
For $p=1$ the matrix $\bold A$ is equal to $-a_{1}$.

\begin{proposition}
If $\{X_t, t\geqslant0\}$ is independent of the L\'evy process $\{L_t, t\geqslant0\}$ where $E[L_{1}^2]<\infty$ then $\{X_t, t\geqslant0\}$ is a
second-order stationary process if and only if the eigenvalues $\lambda_i$, $i=1,2,\ldots, p$ of the matrix $\bf A$ all have negative real parts.
\end{proposition}

\begin{remark}
The eigenvalues of the matrix $\bf A$ are the same as the zeroes of autoregressive polynomial $a(z)$.
\end{remark}

Under the specific condition on the eigenvalues $\{Y_t, t\geqslant0\}$ is a causal function of $\{L_t, t\geqslant0\}$. So $Y_t={\bf b}^{\prime}X_t$
have the following moving average representation
\begin{align}
Y_{t}=\int_{-\infty}^{t}{\bf b}'e^{{\bf A}(t-u)}{\bf e}\, dL_{u}.
\end{align}
and equivalently $Y_{t}=-\int_{0}^{\infty}{\bf b}'e^{{\bf A}u}{\bf e}\, dL_{(t-u)}$. Also, under this condition and distinct eigenvalues and by the
fact that CARMA process $\lbrace Y_{t},\, t\geq 0\rbrace $ can be written as a linear combination of some continuous-time autoregressive process of
order one driven by L\'evy, the background driving L\'evy process is recoverable and the kernel can be written as
$${\bf b}'e^{{\bf A}u}{\bf e}=\sum_{i=1}^{p}\frac{b(\lambda_i)}{a'(\lambda_i)}e^{\lambda_iu}I_{(0, \infty)}(u)$$
where $a^{\prime}(\cdot)$ is the first derivative of the polynomial $a(\cdot)$.
Thus the above equations provide a more general definition of CARMA processes in terms of the Levy process. We used a sampled process in equally space
to estimate the parameters of the underlying continuous-time processes. The non-decreasing property of the driving L\'evy process and the
non-negativity of the corresponding discrete-time increments permits and an efficiency estimation procedure.

\subsection{Affine term structure}
In an ATS model, the interest rates and some derivatives at any point in time are time invariant. This model has flexible structure for controlling
the dynamics of default risk free bonds \cite{Bjork}. ATS model starts from the assumption that there are no arbitrage opportunities in financial
markets and implies the existence of a strictly positive stochastic process.
\begin{definition}
If the term structure $ F $ has the form
\begin{align}
P(t,T)=F(t,r_{t},T)
\end{align}

\noindent where
\begin{align}\label{1.0}
F(t,X,T)=e^{A(t,T)-B(t,T)X}
\end{align}

\noindent and A,B are deterministic functions, then the model is said to possess an ATS. The function A and B are two real variables
function of t and T but conceptually it is easier to think of A and B as being functions of t, while T serves as a parameter.
\end{definition}
It turns out that the existence of an ATS is extremely pleasing from an analytical and a computational point of view. So we are interested to
understand when such a structure appears.

\subsection{Credit default swap}
CDS is a derivative which the documentation identifies reference entity or reference obligation. The reference entity is the issuer of the debt
instrument and can be a corporation, a sovereign government or a bank load. When there is a reference entity, the party to the CDS has an option to
deliver one of the issuer's obligation subject to pre-specified constraints \cite{fabozzi}. \\
In a single name CDS, B agrees to pay the default payment to A if a default has happened. If there is no default of the reference security until the
maturity of the default swap, counter party B pays nothing. A pays a fee for the default protection and the fee can be either a regular fee at interval
until default. If a default occurs between two fee payment dates, A still has to pay the fraction of the next fee payment that has occurs until the
time of default.\\
A premium leg (PL) of a CDS contract is a constant premium (the CDS spread) which has to be paid by the protection buyer at the maturity of the
contract. The opposed default leg (DL) which is the credit event that occurs before maturity has to be reduced by the protection seller, otherwise
there is no payment due. The fair payment is calculated by the following formula.
\begin{align}
PV_{DL}:=E_{Q}[DL(C_{s}^{T})\mid \mathcal{G}_{s}]=E_{Q}[PL(C_{s}^{T}) \mid \mathcal{G}_{s}]=:PV_{PL},
\end{align}
where $ PV $ is the present value, $E_{Q}$  the conditional expected value with respect to measure $\mathcal{G}_{s}$ which is the market information
available up to time $s\geq0$ and also $C_{s}^{T}$ is the premium process which is starting at time $s\in[0, T]$ with maturity $T$. There are several
models that deal with CDS to explore the intensity and survival probability such as structural and reduced form models and both of them are used to
model credit risk.
\begin{definition}\label{d2.1}
(Default time). The default time $\tau:\Omega\rightarrow[0.\infty]$ is a random time representing the time of the credit event of the reference entity.
It's corresponding default process which is denoted by
$H:=\lbrace H_{t}, t \geqslant 0\rbrace:=\lbrace {I_{\lbrace {{\tau\leqslant t}\rbrace}}, t \geqslant 0\rbrace}$.
\end{definition}
We assume that for every $t \geqslant 0 $, $\mathcal{G}_{t}:=\mathcal{F}_{t}\vee \mathcal{H}_{t}=\sigma (\mathcal{F}_{t}\cup \mathcal{H}_{t})$
where  $\Bbb F:=\lbrace \mathcal{F}_{t}\rbrace_{t\geqslant0} $ denotes the filtration representing the default-free,
information and $\lbrace \mathcal{H}_{t}\rbrace_{t\geqslant0}$ denotes the natural filtration of the default process $\Bbb F$, representing the
defaultable, obligor-specific information. Therefore, $\mathcal{G}_{t}$ is econompassing all default-free and defaultable information available to
market participants up to time $t$. As usual, we assume that $\lbrace \mathcal{G}_{t}\rbrace_{t\geqslant0}$ satisfies the conditions of right
continuity and completeness. Note that $\tau$ is a stopping time with respect to
$\lbrace\mathcal{H}_{t}\rbrace_{t\geqslant0}$ and consequently to $\lbrace \mathcal{G}_{t}{\rbrace_ {t\geqslant0}}$, but not with respect to $\Bbb F$.
\begin{definition}\label{def2.1}
Let $(\Omega, \mathcal{F}, P, \Bbb F)$ be a filtered probability space. The $\sigma$-algebra on $\Bbb R^{+}\times \Omega$
generated by all sets of the form $\lbrace  0\rbrace\times A$, $A\in \mathcal{F}_{0}$, and $(a,b] \times A, a<b$ is said to be predictable
$\sigma$-algebra for the filtration $\Bbb F$.
\end{definition}
\begin{definition}
A real-valued process $\{X_{t}, t\in \Bbb R^{+}\}$ is called $predictable $ with respect to filtration  $\Bbb F$, or $\mathcal{F}_{t}$-predictable, if
as a mapping from $ \Bbb R^{+}\times \Omega\rightarrow \Bbb R$ it is measurable with  respect to the
predictable $\sigma-$algebra generated by this filtration.
\end{definition}

\begin{proposition}
Every predictable process is progressively measurable.
\end{proposition}

\begin{definition}
A process $\gamma=\lbrace\gamma_{t}, t\geqslant0 \rbrace$ is called intensity process of default rate of $\tau$, if it is an $\Bbb F$-progressive
process with the following properties for every $t\geqslant0$\\
\noindent
(i) $\gamma_{t}\geq 0$\\
(ii) $\int_{0}^{t} \gamma_{s}ds<\infty$ a.s\\
(iii) for $0\leq s \leq t$, the conditional survival probability $ Q(\tau > t \mid \mathcal{G}_{s})$ is given by
\vspace{-1mm}
\begin{align}
Q(\tau >t\mid \mathcal{G}_{s})=E[I_{\lbrace\tau >t\rbrace}\mid \mathcal{G}_{s}]=I_{\lbrace\tau >t\rbrace}E\Big[exp \Big \lbrace -\int _{s}^{t}\gamma
_{u} du \Big\rbrace \mid \mathcal{F}_{s}\Big].
\end{align}
\end{definition}
In the content of CDS, we apply Key Lemma in \cite{bielechi} to satisfy PL and DL.
\begin{lemma}\label{keylamma}(Key Lemma). Let $X$ be an $\mathcal{F}_{t}$-measurable random variable and \\$Z=\lbrace Z_{t}, {t\geqslant 0}\rbrace$ an
$\Bbb F$-predictable (bounded) process. Then\\
(i) for all $0\leqslant s\leqslant t$, we have
\begin{align}
E[X\,I_{\lbrace\tau >t\rbrace}\mid \mathcal{G}_{s}]=I_{ \lbrace\tau >t\rbrace}E\Big[X\,exp \Big \lbrace -\int _{s}^{t}\gamma _{u} du \Big\rbrace \mid
\mathcal{F}_{s}\Big],
\end{align}
(ii) For all $s\geqslant0$, we have
\begin{align}
E[Z_{\tau}\mid \mathcal{G}_{s}]=I_{ \lbrace\tau \leqslant t\rbrace}Z_{\tau}+I_{ \lbrace\tau >t\rbrace}E\Big[\int_{s}^{\infty}Z_{t}\gamma_{t}exp \Big
\lbrace -\int _{s}^{t}\gamma _{u} du \Big\rbrace dt \mid \mathcal{F}_{s}\Big].
\end{align}
\end{lemma}

\begin{proposition}\label{pro2.4}CDS is fair default if
\begin{align}
PL(C_{s}^{T})=\int_{s}^{s+T}C_{s}^{T}D_{s}^{t}I_{\lbrace \tau >t\rbrace}dt
\end{align}
\vspace{-5mm}
 and
\begin{align}
DL(R_{t})=(1-R_{t})D^{\tau}_{s}I_{\lbrace \tau < s+T\rbrace}
\end{align}
where $C^T_s$ is the premium, $D^t_s$ discount factor, $\tau$ the default time and $R_{t}$ is a stochastic recovery rate function.
\end{proposition}

\section{Affine term structure of CARMA model}
ATS model is often used to give an account of any arbitrage-free model in which bond yields are affine functions of some state vector $x$.
Affine models are a special class of term structure models, which yield $y(T) $ of a $T$-period bond as $y(T)=A(T)+B(T)x$ for coefficients $A(T)$ and
$B(T)$ that depend on maturity $T$.
The functions $A(T)$ and $B(T)$ make these yield equations consistent with each other for different values of $T$ and the state dynamics. The main
advantage of affine models is to have tractable solutions for bond yields which are used because otherwise they are costly computed. The functional
form of bond yields is obtained from computing risk-adjusted expectations of future short rates. Now by the following proposition we show that the
L\'evy-driven CARMA process is ATS.
\begin{proposition}\label{3.1}
ATS of the L\'evy-driven CARMA process $\lbrace X_{t}, t\geqslant 0\rbrace$ by the dynamic $dX_{t}=\bold A X_{t}dt+\bold e\,dL_{t}$ is
\vspace{-3mm}
\begin{align}
P(t,T)=e^{A(t,T)-B(t,T)r(t)}
\end{align}
where
\vspace{-5mm}
\begin{align}\label{00}
A(t,T)&=(\frac{{\bf A}^{-1}\bf e}{2})^2
\big[{\bold A }B^{2}(t,T)-2B(t,T)+2(T-t)\big]
\end{align}
\vspace{-5mm}
and
\begin{align}\label{11}
B(t,T)&={\bf A}^{-1}[{e^{{\bold A} (T-t)}-1}].
\end{align}
\end{proposition}
where the matrix $\bold A$ and the vector $\bold e$ are introduced in subsection 2.1.\\ \\
{\bf Proof}: The ATS systems of equations of CARMA models are
\begin{equation}\label{1}
B_{t}(t,T)+\bold A B(t,T)=-1
\end{equation}
\vspace{-1mm}
and
\vspace{-5mm}
\begin{align}\label{2}
A_{t}(t,T)=-\dfrac{1}{2}{\bf e}^{2}B^{2}(t,T)
\end{align}
with initial values $A(T,T)=0$ and $B(T,T)=0$.
For fixed $T$, the system (\ref{1}) is a simple linear first order differential equation, so we get to the equation $B(t,T)$ in (\ref{11}).
To verify the correctness, we observe that $B_{t}(t,T) =-e^{\bold A (T-t)} $ and by replacing it in (\ref{1}) we have
$$
B_{t}(t,T)+\bold A B(t,T)=-e^{\bold A (T-t)}+\bold A \big({\bold A }^{-1}[e^{\bold A (T-t)}-1]\big)=-e^{\bold A (T-t)}+e^{\bold A (T-t)}-1=-1
$$
Now we find $ A(t,T) $ and for this, we get integral from (\ref{2}) as
$$A(T,T)-A(t,T)=\int_{t}^{T}\dfrac{-\bold e^{2}}{2}B^{2}(s,T)ds.$$
Therefore
\begin{align*}
A(t,T)&=\frac{({\bold A ^{-1}\bold e})^2}{2}\int_t^T[e^{\bold A (T-s)}-1]^{2}ds
=\frac{({\bold A ^{-1}\bold e})^2}{2}\int_t^T[e^{2\bold A (T-s)}-2e^{\bold A (T-s)}+1]ds\\
&=\frac{({\bold A ^{-1}\bold e})^2}{2}\big[\frac{1}{2}{\bold A ^{-1}}e^{2\bold A (T-t)}-2{\bold A ^{-1}}e^{\bold A (T-t)}+\frac{3}{2}{\bold A
^{-1}}+(T-t)\big]\\
&=\frac{({\bold A ^{-1}\bold e})^2}{2}\Big[\frac{{\bold A ^{-1}}}{2}[e^{2\bold A (T-t)}-2e^{\bold A (T-t)}+1]+\big[(T-t)-{\bold A ^{-1}}[e^{\bold A
(T-t)}-1]\big]\Big],
\end{align*}
and by replacing the equation $B(t,T)$ we get to equation (\ref{00}).
So we can write the CARMA model as an ATS model.

\section{Premium leg under stochastic recovery rate}
Many companies try to reduce the risks that they encounter daily. So it is important to decrease the risk of large losses and to increase a financial
firm's resilience. One factor that determines the extent of losses is the recovery rate on loans and bonds that are in default. Financial companies and
researchers commonly assume that the recovery rate is constant but in
practice actual recovery rates vary with respect to time. This assumption is important because additional risk is introduced when the recovery rate is
not
constant. We develop the credit risk models by considering the stochastic recovery rate in the purpose of profitably pricing the CDS premium leg.
Before studying the CDSP-SRR model we give a description of CDSP-CRR model.\\ \\
The constant recovery rate of a reference entity is the fraction of the notional debt outstanding by the reference entity and is to be recoverable
after the default event. The constant parameter is denoted by $R$ which is between 0 and 1.
In CDS contracts one fixes this parameter with respect to amount of contract.
The loss compensation payment of amounts is $1-R$. By denoting the random time of the credit event $\tau$ and the random cash flows of the swap legs'
the CDS price is provided \cite{Eifert}. Since the parameters in real market are stochastic, they affect on each other and estimate by a variety of
conditions. It is essential to be aware of the feature of the parameters and try to choose them in the right way.\\ \\
Now we assume that the recovery rate is stochastic and depends on the default intensity $\gamma_{t}$ and then find a model for it. This should fulfill
three necessary properties.
First, the domain of the stochastic recovery rate is on $\Bbb R^{+}$. The second crucial condition is, in order to ensure a negative correlation of
default probabilities with recovery rates, we need to use a function that has a negative or non-positive first derivative. It is known that realized
recovery
rates are negatively related to aggregate default rates. Therefore, we impose the same condition on implied recovery rates. So we have $R_{t}=\beta
_{2}+\beta _{0}e^{\beta_{1}\gamma _{t}}$ which is between 0 and 1. A convenient possibility is an exponential function where by considering the above
properties we have $\beta _{0},\beta _{2}\in(0,1), \beta _{1}\leq 0$.
 Now by these assumptions we present an intensity based representation of CDSP-SRR model and then find an innovative model for CDS default leg.

\begin{proposition}
Let r be the short rate and its corresponding discount factor $D_{s}^{t}$ be $\Bbb F$-predictable then the fair CDS default has the following
intensity-based expression
\begin{align}\label{4.1}
C^{T}_{s} =\dfrac{E\Big [\int_{s}^{s+T} (1-R_{t})D_{s}^{t} \gamma_{t} exp\lbrace -\int_{s}^{t}\gamma _{u} du\rbrace dt\mid\mathcal{F}_{s}\Big]}{E\Big
[\int_{s}^{s+T} D_{s}^{t} exp\lbrace -\int_{s}^{t}\gamma _{u} du\rbrace dt\mid \mathcal{F}_{s}\Big]},\,\,\,\,\,s<t.
\end{align}
\end{proposition}
Proof: By applying Lemma \ref{keylamma} and using the result of Bielecki et al. \cite{bielechi} we have
\begin{align*}
 PV_{PL}(C^{T}_{s})&=E\Big [\int_{s}^{s+T} D_{s}^{t} I_{\lbrace \tau >t \rbrace} C^{T}_{s} dt\mid\mathcal{G}_{s}\Big]\\&=C^{T}_{s} I_{\lbrace \tau
 >s\rbrace} E\Big [\int_{s}^{s+T} D_{s}^{t} exp\lbrace -\int_{s}^{t}\gamma _{u} du\rbrace dt\mid \mathcal{F}_{s}\Big]\\
PV_{DL}(R_{t})&=E\Big [ (1-R _{t})D_{s}^{\tau}  I_{\lbrace\tau <s+T\rbrace }\mid \mathcal{G}_{s}\Big]\\&= I_{\lbrace\tau >s\rbrace}E\Big
[\int_{s}^{s+T}
(1-R _{t})D_{s}^{t} \gamma_{t} exp\lbrace -\int_{s}^{t}\gamma_{u} du\rbrace dt\mid \mathcal{F}_{s}\Big].
\end{align*}
Therefore, equating both present values in the above equation we get to the result.

\begin{proposition}{(Credit triangle)} We assume that the continuously paid premium $C^{T}_{s}$ is not fixed at the starting date $s\geq0$ and it is
time-varying during the contractual tenor $[s,s+T]$, named floating premium. Then the CDS default $\gamma^{T} $ is equal for all tenors T flat spread,
$C^{T}\equiv C$ and
\begin{align}
C^{T}_{s}=(1-R_{s})\gamma_{s},
\end{align}
where the stochastic recovery rate $R_{s}$ is $\beta _{2}+\beta _{0}e^{\beta_{1}\gamma _{s}}$.
\end{proposition}
Observations are made on a daily basis, however punctual gaps may occur in our data,
so for $0<s<t, s:=t-1$, we model the log-returns of the CDS premia equivalently to the log-returns of the default rate process by the followings
\begin{align}\label{4.3}
C^T_t=(1-R_t)\gamma_t=(1-\beta_2-\beta_0e^{\beta_{1}\gamma_{t}})\gamma_t,
\end{align}
therefore
\begin{align}\label{4.4}
logC^T_t-logC^T_s=log\Big(\dfrac{1-R_t}{1-R_s}\Big)+log\Big(\dfrac{\gamma_t}{\gamma_s}\Big)=\int^{t}_{s}Y_{u}du
\end{align}
where $\{Y_t, t\geqslant0\}$ is the L\'evy-driven CARMA process as the log-return which is described in Section 2.

\section{Simulation Study and Real Data}
In this section, the CARMA$(p,q)$ processes with different orders are simulated. We show that how to simulate the CARMA$(p,q)$. We explain the
schematic representation of the model by compound Poisson process with normal distributed jumps in subsection 5.1. In subsection 5.2, we use a real
data from entity companies (242 firms) to compare the CDSP-CRR model and the introduced CDSP-SRR model. In this case we apply the log-likelihood and
BIC for comparing these models.

\subsection{Simulation}
We evaluate the performance of CDSP-SRR by using CAR(1), CAR(2) and CARMA(2,1) processes. We
modify the {\bf Yuima} package \cite{Mercuri} corresponding to the presented models and describe the required steps as the following.\\ \\
1. As a default in $ \bf simulation$ function a continuous process is sampled at equally spaced time instants   $0, h, 2h, \ldots, Nh$ where $N$ is the
number of observations and $h$ is the step length. In this case, $h=1$, $N=3000$ and daily time is required. We set an initial value for the parameters
of CARMA model processes. Also, we use moving averages of length 5 to handle the missing data.\\
\\
2. In the introduced model, we are to estimate the parameters that govern the implied recovery rate $R_{t}$ in this step. For the model with stochastic
recovery rate, we estimate the parameters $\beta_{0},\beta_{1}$ and $\beta_{2}$ while for the model with constant recovery rate just we set
$\beta_{0}=R$ as a constant parameter. The Markov Chain Mont Carlo method allows a straight forward calculation of $95\%$ confidence interval, to
determine credible intervals for $\beta$ parameters.\\
\\
3. We use quasi-maximum likelihood estimation method for estimating the parameters of the CARMA model. The arguments in the $\bf qmle$ function provide
the new $\bf Yuima$ function $\bf Carma.Noise$ with estimated L\'evy increments.\\
\\
4. Finally we compare BIC of the CDSP-CRR and the CDSP-SRR models to select the suitable model for CDS premia. We implement the simulation technique
through the following simulation example and show the results in the Tables 1 and 2.\\

\noindent Following the simulation steps we illustrate the CARMA(2,1) processes under CDSP-CRR and CDSP-SRR models in Figures 1 and 2. It is shown that
the CDSP-SRR model because of the time-varying recovery rate has more fluctuation than the others.\\

\begin{example}
We simulate the sample path of CAR(1) with coefficients $a_{1} =6$ as $(D + a1=6)Y_{t} = DL_{t}$. We choose $\beta_{0}=0.378$, $\beta_{1}=-0.0095$ and
$\beta_{2}=0.637$.
The simulations of the CDSP-CRR and CDSP-SRR models are plotted in Figures 1 and 2 and the results in comparing them are recorded in Table 1.
\end{example}

\input{epsf}
\epsfxsize=7in \epsfysize=2in
\begin{figure}\vspace{-.1in}
\centerline{\hspace{-.2in}\epsfxsize=5in \epsfysize=2.5in
\epsffile{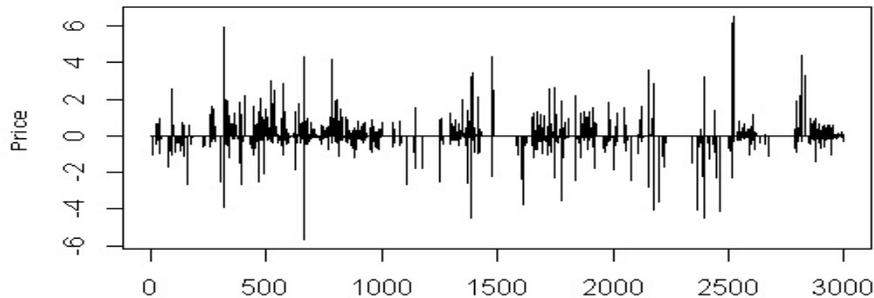}}\vspace{-.5in}
\caption{\footnotesize{ Simulation of CARMA(2,1) under CDSP-CRR model.}}
\end{figure}

\input{epsf}
\epsfxsize=7in \epsfysize=2in
\begin{figure}\vspace{-.1in}
\centerline{\hspace{-.2in}\epsfxsize=5in \epsfysize=2.5in
\epsffile{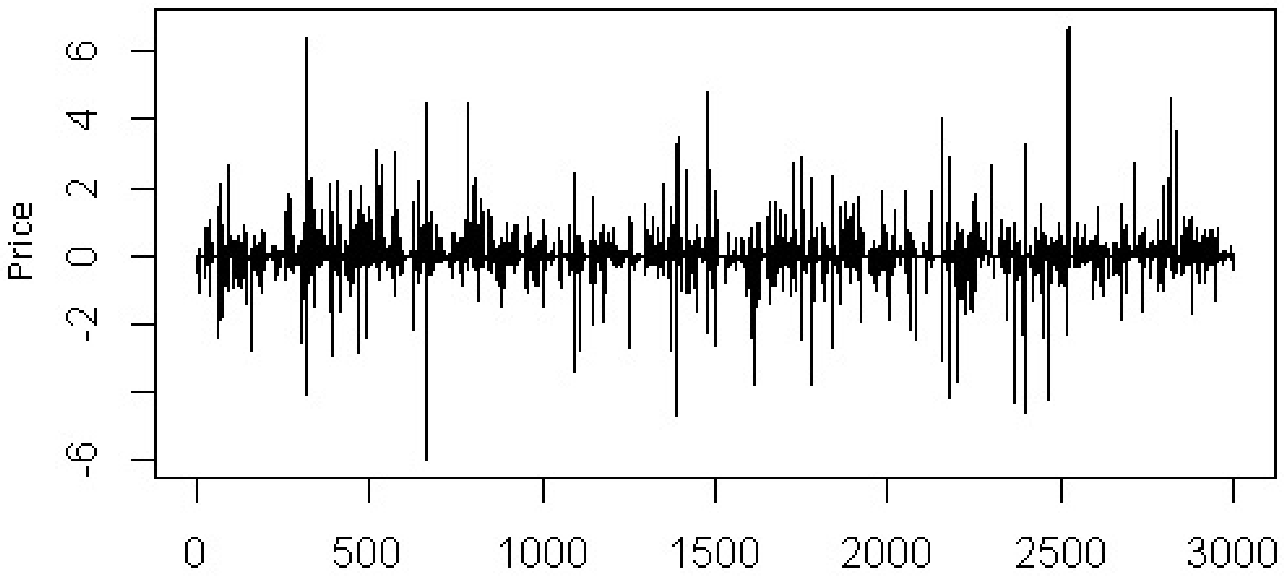}}\vspace{-.6in}
\caption{\footnotesize{ Simulation of CARMA(2,1) under CDSP-SRR model.}}
\vspace{-2mm}
\end{figure}

\begin{table}[h!]
\begin{center}
\small
\begin{tabular}{|c c c c c|}
\hline
Model &\vline& BIC &\vline& log-likelihood\\ [0.1ex]
\hline\hline
CDSP-SRR &\vline &\scriptsize 2123.2826  &\vline& \scriptsize -1033.6413\\
\hline
CDSP-CRR &\vline &\scriptsize 4562.5487 &\vline& \scriptsize  -2269.2648\\
\hline
\end{tabular}
\caption{\footnotesize BIC and log-likelihood of CAR(1) with CDSP-CRR and CDSP-SRR models.}
\label{table:1}
\end{center}
\end{table}
\vspace{-6mm}

\begin{example} According to the simulation, we generate a set of sample path of \\CARMA(2,1) with parameters $a_{1} = 1.39631$, $ a_{2} = 0.05029$,
$b_{0} = 1$ and $b_{1}=2$. The corresponding stochastic differential equation of the process is $(D_{2} + 1.39631D + 0.05029)Y_{t} = (2 + D)DL_{t}.$
We choose appropriate $\beta_{0}=0.0378$, $\beta_{1}=-0.0095$ and $\beta_{2}=0.637$. Then, we generate the data based on CDSP-CRR and CDSP-SRR models
which are defined in (\ref{4.4}). The results of BIC and log-likelihood values are recorded in Table 2.
\end{example}

\begin{table}[h!]
\begin{center}
\small
\begin{tabular}{|c c c c c|}
\hline
Model &\vline& BIC &\vline& log-likelihood\\ [0.1ex]
\hline\hline
CDSP-SRR &\vline &\scriptsize 4801.2438  &\vline& \scriptsize -2380.6068\\
\hline
CDSP-CRR &\vline &\scriptsize 8740.4511 &\vline& \scriptsize  -4350.2096\\
\hline
\end{tabular}
\caption{\footnotesize BIC and log-likelihood of CARMA(2,1) with CDSP-SRR and CDSP-CRR models.}
\label{table:1}
\end{center}
\end{table}
\vspace{-4mm}
\noindent These differences can affect the price of the CDS premia in a real market and all companies are affected by the stochastic recovery rate.

\subsection{Real Data}
To compare the performance of the proposed model, we consider 5-years CDS spread (T=5) observed daily between January 2002 until November 2012 (2829
trading days) for N=242 firms. The reference entities are from European and North American companies including different sectors such as Banks,
Electric power and other financial companies. For any company we have different staring dates $t_{min}$ and ending date $t_{max}$ by maturity time
$T={t_{min},t_{min}+h,t_{min}+2h,...,t_{max}}$ with uniform step size $h$.\\
Our model is based on the observed (one-period) log-returns denoting the discrete-time observations defined by (\ref{4.4}). Figures 1 and 2 reveal that
the ING bank of the real data, clarifies the effects of CDSP-SRR model.  By applying the CDSP-CRR and CDSP-SRR models to every credit entity of CDS we
fit CARMA models to data and then estimate the parameters. Also, we compare some of the companies in terms of the BIC in Table 3. In this table, we show that the CDSP-SRR model has smaller BIC either than the CDSP-CRR model.

\input{epsf}
\epsfxsize=7in \epsfysize=2in
\begin{figure}\vspace{-3cm}
\centerline{\hspace{-.2in}\epsfxsize=5in \epsfysize=2.5in
\epsffile{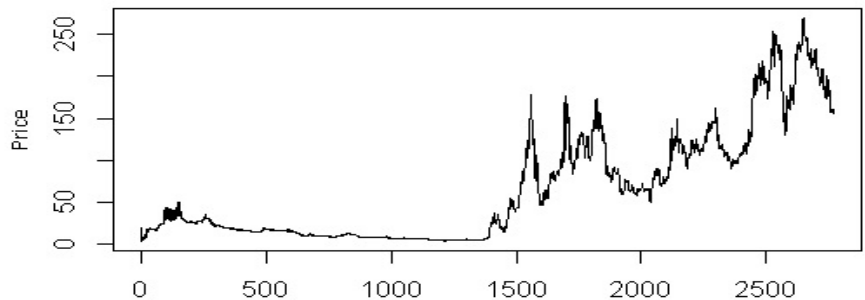}}\vspace{-.5in}
\caption{\footnotesize Realization of ING bank data.}
\end{figure}

\input{epsf}
\epsfxsize=7in \epsfysize=2in
\begin{figure}\vspace{-.1in}
\centerline{\hspace{-.2in}\epsfxsize=5in \epsfysize=2.5in
\epsffile{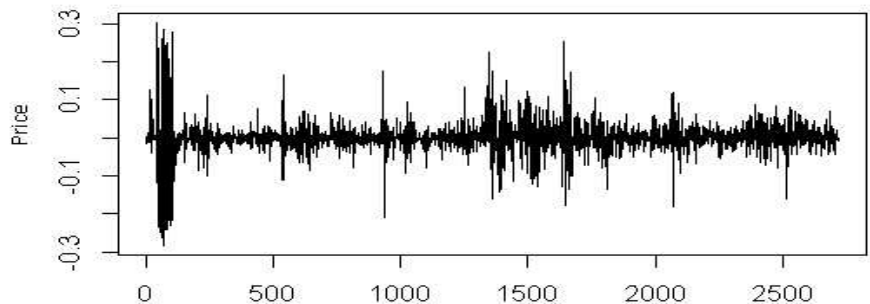}}\vspace{-.5in}
\caption{\footnotesize Fitting the ING bank data with CAR(1) process.}
\end{figure}

%\vspace{5mm}
\begin{table}[h!]
\begin{center}
\small
\begin{tabular}{|c c c c c|}
\hline
Companies &\vline&  CDSP-SRR model  &\vline& CDSP-CRR model \\ [0.1ex]
\hline\hline
AMERICANEXPRESS &\vline &\scriptsize -10137.9672 &\vline& \scriptsize -9228.8668 \\
\hline
BMW &\vline &\scriptsize -11448.7974 &\vline& \scriptsize -10104.0573 \\
\hline
SANO &\vline &\scriptsize -12684.5042 &\vline& \scriptsize -11175.8189 \\
\hline
KPN&  \vline &\scriptsize -12889.6028 &\vline& \scriptsize -11596.5878 \\
\hline
DAIMLER&\vline &\scriptsize -11892.0160 &\vline &\scriptsize -10742.6483\\
\hline
ING&\vline &\scriptsize -8114.2239 &\vline &\scriptsize -7088.9918\\
\hline
DEUTSCHE&\vline &\scriptsize -11218.7720 &\vline &\scriptsize -9945.8812\\
\hline
CONOCO PHILLIPS&\vline &\scriptsize -11113.9200 &\vline &\scriptsize -9626.3721\\
\hline
WAL-MART&\vline &\scriptsize -10856.7632 &\vline &\scriptsize -9825.9177\\
\hline
MCDONALDS&\vline &\scriptsize -11676.4504 &\vline &\scriptsize -10211.3201\\
\hline
COMCAST&\vline &\scriptsize -5211.7253 &\vline &\scriptsize -4661.1184\\
\hline
\end{tabular}
\caption{\footnotesize Comparing model performance with BIC}
\label{table:1}
\end{center}
\end{table}

\section*{Conclusion}
To modeling the financial time series data such as CDS spread, we review the introduced model and generalized it. Considering the fact that the
recovery rate of a CDS is not constant, we improve this model by considering the stochastic recovery rate. This assumption makes a better and more
flexible model than the previous one. A continuous-time ARMA process driven by L\'evy which is stationary is used to fit the premium of a CDS.
According to the log-likelihood value and BIC, we get the results that CDS-SRR model has better performance for NIG distribution of the jumps.
The empirical studies of this paper is based on the CDS spread of some companies use CAR(1), CAR(2) and CARMA(2,1). However, there are several
limitations in the model, besides estimation of the CARMA model, the results are better and it can be better if we set the real plausible amount of
recovery rate in our model. Also, by changing the structure of the stochastic recovery rate we can improve the efficiency of the model.

\end{document}